\documentclass[a4paper,preprint,superscriptaddress,preprintnumbers,nofootinbib]{revtex4}
\usepackage{amsmath,amssymb}
\usepackage{bm}
\usepackage[dvips]{graphicx}

\begin{document}

\title{Uniqueness and Significance of Weak Solution of 
Non-perturbative Renormalization Group Equation \\
to Analyze Dynamical Chiral Symmetry Breaking
\footnote{Presented at ``SCGT12 KMI-GCOE Workshop on Strong Coupling Gauge
Theories in the LHC Perspective'', 4-7 Dec. 2012, Nagoya University.}}

\author{Ken-Ichi \surname{Aoki}}
\email{aoki@hep.s.kanazawa-u.ac.jp}
\affiliation{Institute for Theoretical Physics, Kanazawa University, Kanazawa 920-1192, Japan}

\author{Shin-Ichiro \surname{Kumamoto}}
\email{kumamoto@hep.s.kanazawa-u.ac.jp}
\affiliation{Institute for Theoretical Physics, Kanazawa University, Kanazawa 920-1192, Japan}

\author{Daisuke \surname{Sato}}
\email{satodai@hep.s.kanazawa-u.ac.jp}
\affiliation{Institute for Theoretical Physics, Kanazawa University, Kanazawa 920-1192, Japan}

\preprint{KANAZAWA-13-05}

\begin{abstract}
We propose quite a new method of analyzing the dynamical chiral symmetry
breaking in gauge theories. Starting with the non-perturbative renormalization 
group equation for the Wilsonian fermion potential, we define the weak solution
of it in order to mathematically authorize solutions with singularity.
The weak solution is obtained uniquely and it successfully predicts the physically
correct vacuum, chiral condensates, dynamical mass, through its 
auto-convexizing power for the effective potential. 
Thus it works perfectly even for the
first order phase transition in the finite density QCD.  
\end{abstract}
\maketitle

\section{Introduction}

Among various methods to analyze the dynamical chiral symmetry breaking, 
the non-perturbative renormalization group is 
quite effective since it may include non-ladder diagrams which cure the
gauge invariance problem\cite{Sato13}.
In the lowest order approximation, 
the Wilsonian effective action is expressed by a scale dependent 
fermion potential $V_{\rm W}(x,t)$, 
where $x=\int d^4x \bar\psi(x)\psi(x)$ is an operator variable,
$t=\log(\Lambda_0/\Lambda(t))$ is the renormalization scale.
This potential keeps all information of the fermionic interactions, 
the free energy $V_0$, the dynamical mass $M_d$, the 4-fermion interactions G 
{\sl etc} as follows:
\begin{equation}
V_{\rm W}(x,t)= V_0(t) + M_d(t) x + G(t) \frac{x^2}{2} + \cdots .
\end{equation}
The renormalization group equation for $G$ (in case $M=0$) is written as
\begin{equation}
\frac{dG}{dt}= -2 G + \frac{1}{2\pi^2}G^2 ,
\end{equation}
which makes $G$ diverge at a finite scale $t_{\rm c}$ for upper critical initial value. 
This blowup nature 
itself is a correct behavior since it expresses the divergence of the 
susceptibility due to the spontaneous chiral symmetry breakdown\cite{Aoki96, Aoki99}.
However, due to this divergence we can not go beyond $t_{\rm c}$, 
and there is no way to calculate infrared physical quantities 
such as the chiral condensate or the dynamical mass.

\section{Partial differential equation and its weak solution}

Various methods have been used to bypass this singularity, e.g., 
the bare mass{\cite{Miya09}}, auxiliary fields\cite{Aoki96, Aoki99, Gies02}, {\sl etc.} 
Here we propose a new direct
method of solving the renormalization group equation
as a partial differential equation (PDE)\cite{Kuma13}: 
\begin{equation} 
\frac{\partial M(x,t) }{\partial t} +   
\frac{\partial f(M(x,t),t)}{\partial x}  =0,\ 
M=\frac{\partial V_{\rm W}}{\partial x},\ 
f(M,t)=-\frac{1}{4 \pi^2}e^{-4t} \ln [1+M^2 e^{2t}], 
\end{equation}
where $M(x,t)$ is called the mass function.
The 4-fermion interaction $G$ corresponds to
$\frac{\partial M}{\partial x}$ at the origin, and its divergence 
at finite $t_{\rm c }$ means that the above PDE has no
global classical solution beyond $t_{\rm c}$.

\begin{figure}[htbp]
  \begin{center}
    \begin{tabular}{c}
      \begin{minipage}{0.5\hsize}
        \begin{center}
          \includegraphics[clip, width=7cm]{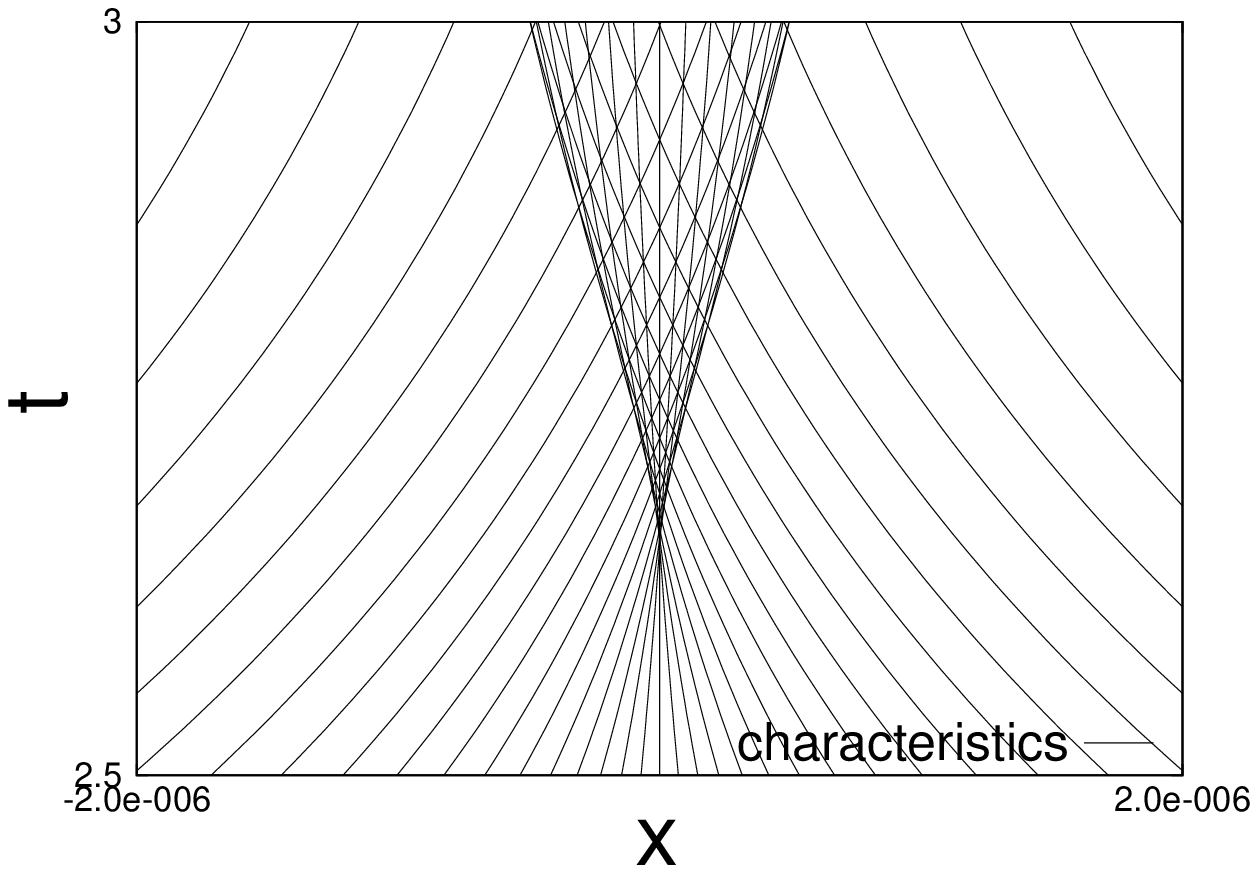}
          \hspace{1cm} (a) 
        \end{center}
      \end{minipage}
      \begin{minipage}{0.5\hsize}
        \begin{center}
          \includegraphics[clip, width=7cm]{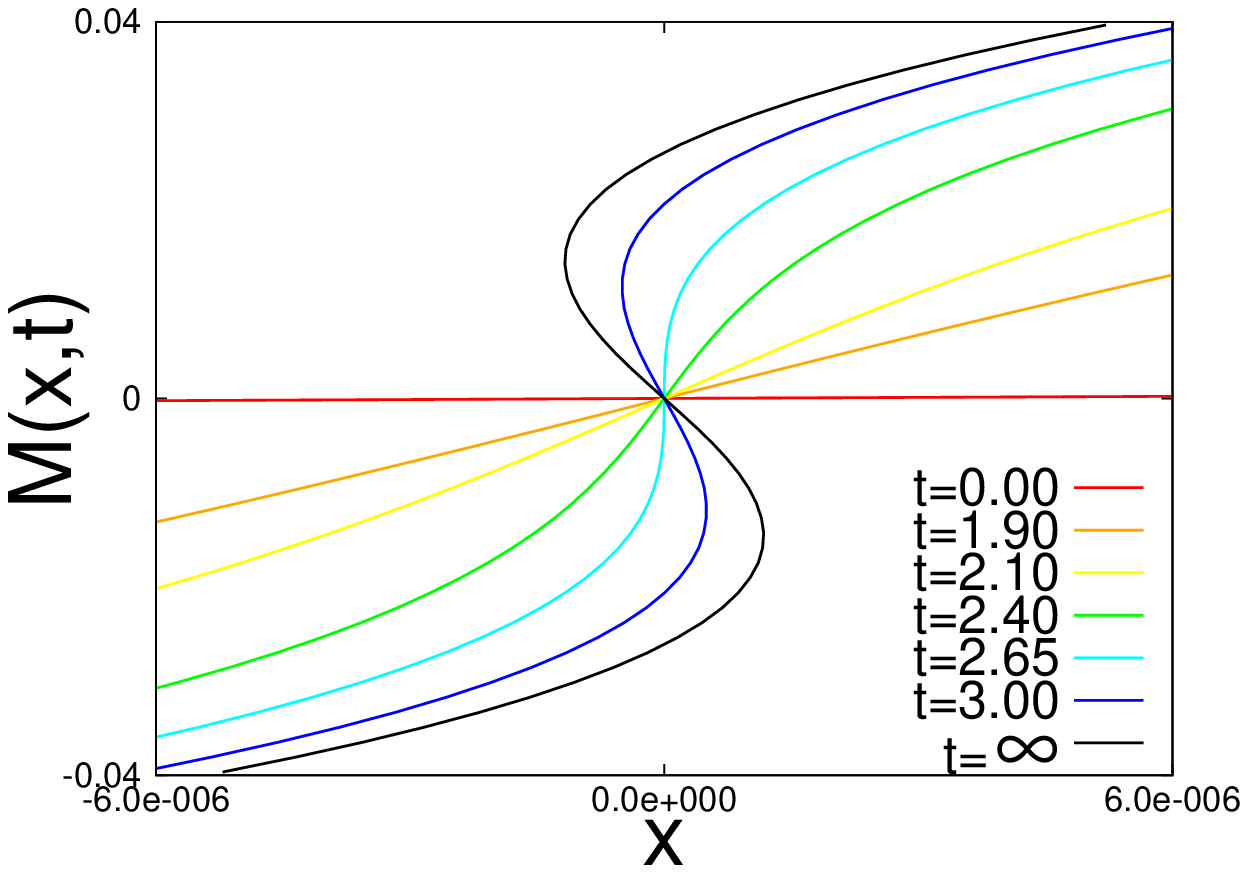}
          \hspace{1cm} (b) 
        \end{center}
      \end{minipage}
    \end{tabular}
\vskip 0.3cm
    \begin{tabular}{c}
      \begin{minipage}{0.5\hsize}
        \begin{center}
          \includegraphics[clip, width=7cm]{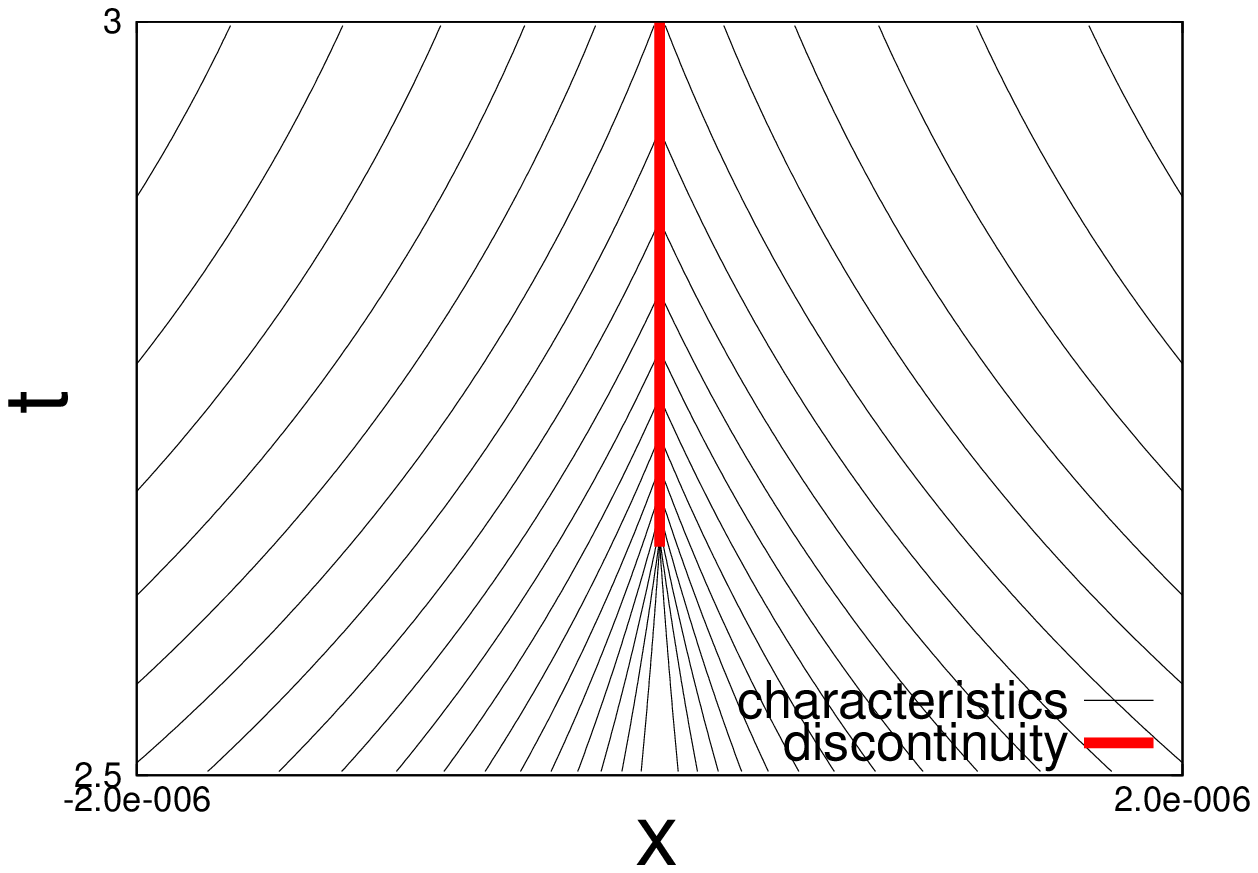}
          \hspace{1cm} (c) 
        \end{center}
      \end{minipage}
      \begin{minipage}{0.5\hsize}
        \begin{center}
          \includegraphics[clip, width=7cm]{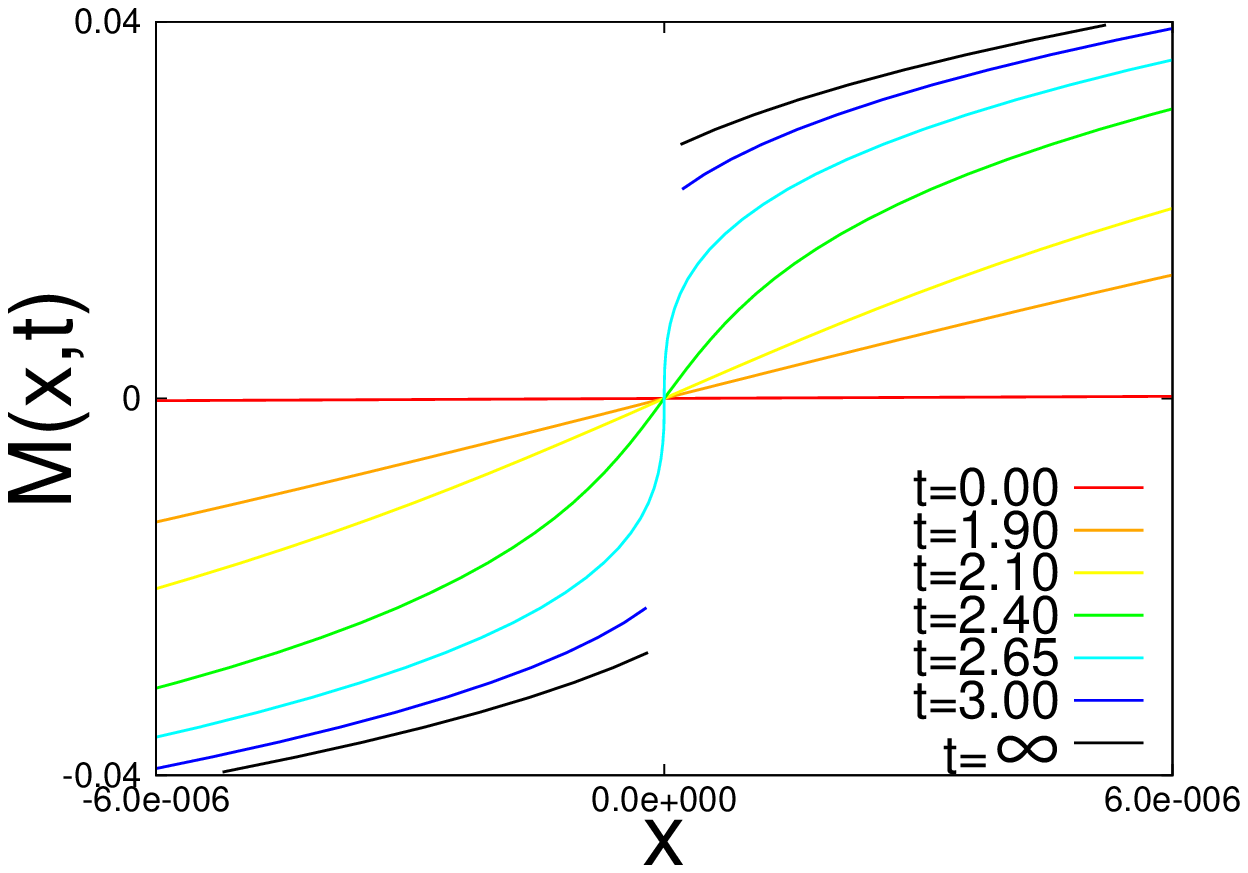}
          \hspace{1cm} (d) 
        \end{center}
      \end{minipage}
    \end{tabular}
    \caption{NJL plot. (a) Characteristics. (b) Mass function. 
(c) Discontinuity. (d) Weak solution.}    
    
    \label{fig:fig1}
  \end{center}
\end{figure}

The expected behavior of physically meaningful solution $M(x,t)$ is plotted in
Fig.~1(d), where $M$ has a jump at the origin after $t_{\rm c}$. 
In order to accommodate such singular solution with discontinuity, we 
write down the weak version of the PDE\cite{Kuma13, Eva01}, 
\begin{equation}  
\int_{0}^{\infty}  dt   \int_{- \infty}^{\infty}  dx   \left[ M  \frac{\partial  \varphi}{\partial t} +f(M,t) \frac{\partial  \varphi}{\partial x} \right] +  \int_{- \infty}^{\infty}  dx  ~M(x,0) \varphi (x,0) =0.
\end{equation}
The weak solution
\footnote{The authors greatly appreciate helpful comments by Prof. Akitaka Matsumura who told us how to construct the weak solution.}
 is defined as to satisfy the above equation for
any smooth and bounded test function $\varphi(x,t)$. 
The weak solution satisfy original PDE except for discontinuity
and position of discontinuity $x=S(t)$ is controlled by the
Rankine-Hugoniot (RH) condition,
\begin{equation} 
\frac{d S(t)}{dt} [M_{+} - M_{-} ] =f(M_{+}) - f(M_{-}),
\end{equation}
where $M_+, M_-$ are right and left limit at the discontinuity point respectively.

To obtain the weak solution, first we set up characteristic curves $x=X(t)$ representing 
the contour lines of $M$, which satisfies
\begin{equation}
\frac{d X(t)}{dt} = \frac{ \partial f(M,t)} {\partial M}.
\end{equation}
Where these curves are crossing each other, 
we must pick up one curve and introduce discontinuity
according to the RH condition, and then we get the unique function $M(x,t)$. 
Fig.~1 shows the Nambu-Jona-Lassinio (NJL) model example 
($g=1.005 g_{\rm c}$) of these procedures. 
The contour lines are plotted in (a), where at some finite $t_{\rm c}$ they 
start crossing with each other. The cross section of the total contours is seen in (b), 
where the derivative $\frac{\partial M}{\partial x}$ diverges 
at the origin at some $t_{\rm c}$.

Note that in obtaining $M(x,t)$ as in (b) there is nothing singular, and it is just a 
motion of `string'. However, to get the renormalized potential we have to
define $M(x,t)$ as a unique function of $x$. Then the RH condition determines 
the discontinuity as shown in (c), and 
the mass function in (d). This is the weak solution of the
PDE and it defines global $t$ solution uniquely.

\section{Weak solution results for the physical quantities}

We show results in the finite density NJL where the first order
phase transition occurs. Contours and mass function are plotted in 
Fig.~2, where at the central region five-fold structure appears
corresponding to the three-fold local minima.
In the renormalization procedure, two discontinuities appear pairwisely,
move towards the origin, and finally merge into
one at the origin as shown in Fig.~2(a).

\begin{figure}[htbp]
  \begin{center}
    \begin{tabular}{c}
      \begin{minipage}{0.5\hsize}
        \begin{center}
          \includegraphics[clip, width=7cm]{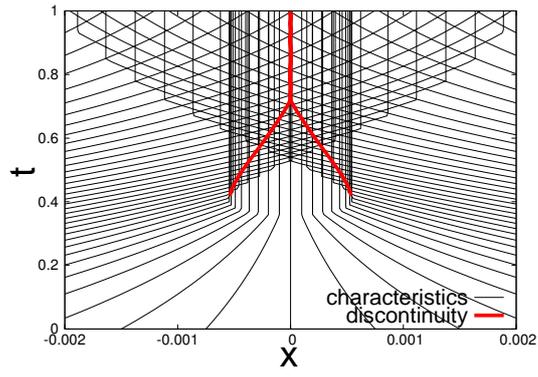}
          \hspace{1cm} (a) 
        \end{center}
      \end{minipage}
      \begin{minipage}{0.5\hsize}
        \begin{center}
          \includegraphics[clip, width=7cm]{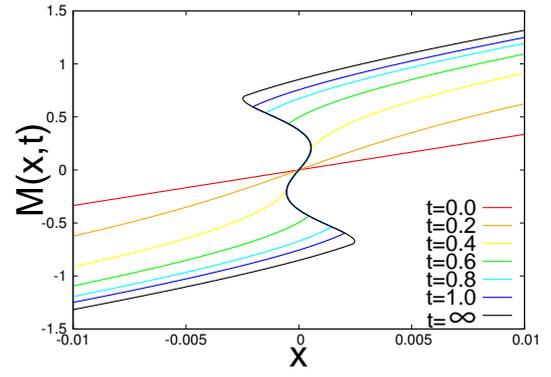}
          \hspace{1cm} (b) 
        \end{center}
      \end{minipage}
    \end{tabular}    
    \caption{NJL $g=1.7 g_{\rm c},~\mu=0.7$. (a) Characteristics and jumps. (b) Evolution of mass function.}    
    
    \label{fig:fig1}
  \end{center}
\end{figure}

Snapshots in the course of renormalization are shown in Fig.~3, where
the mass function $M(t,x)$, the Wilsonian fermion potential $V_{\rm W}$ and 
the Legendre effective potential for $\sigma$ are plotted.
It is astonishing that our method of weak solution uniquely determines 
their singularity structures and the resultant Legendre effective potential 
is always convexized. This means the dynamical mass and the chiral condensates
are uniquely calculated, and perfectly correct in the sense that 
even in case there are multi local minima, the lowest free energy 
minimum is always chosen automatically. This feature is quite a new
finding and shows powerfulness of the purely fermionic non-perturbative renormalization 
group and its weak solution\cite{Kuma13}. This analysis has been applied to
QCD, even with finite density or non-ladder, and proved to work
perfectly to give physical quantities without any ambiguity\cite{Sato13}.

\begin{figure}[htbp]
  \begin{center}
    \begin{tabular}{c}   
      \begin{minipage}{0.3\hsize}
        \begin{center}
          \includegraphics[clip, width=5cm]{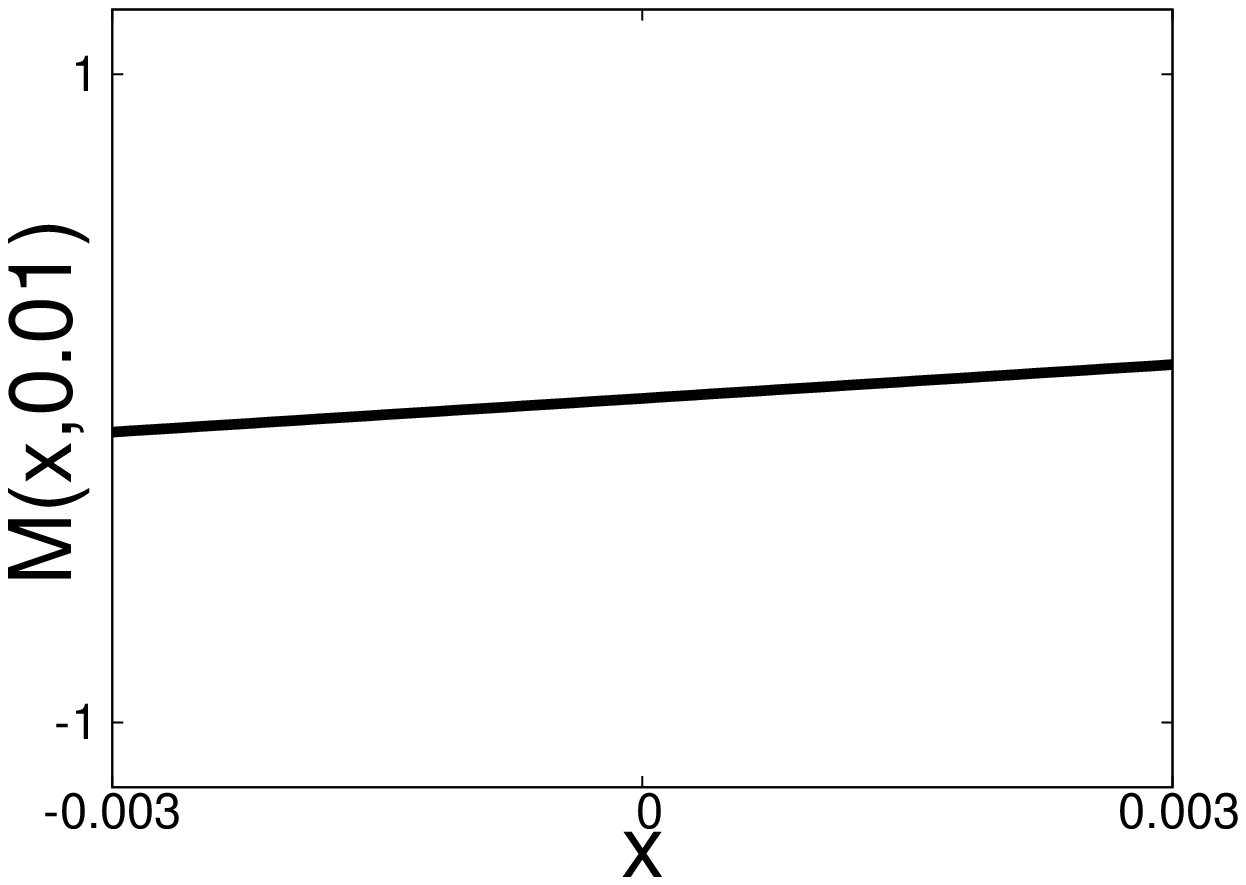}
          \hspace{1cm} (1a) 
        \end{center}
      \end{minipage}
      \begin{minipage}{0.3\hsize}
        \begin{center}
          \includegraphics[clip, width=5cm]{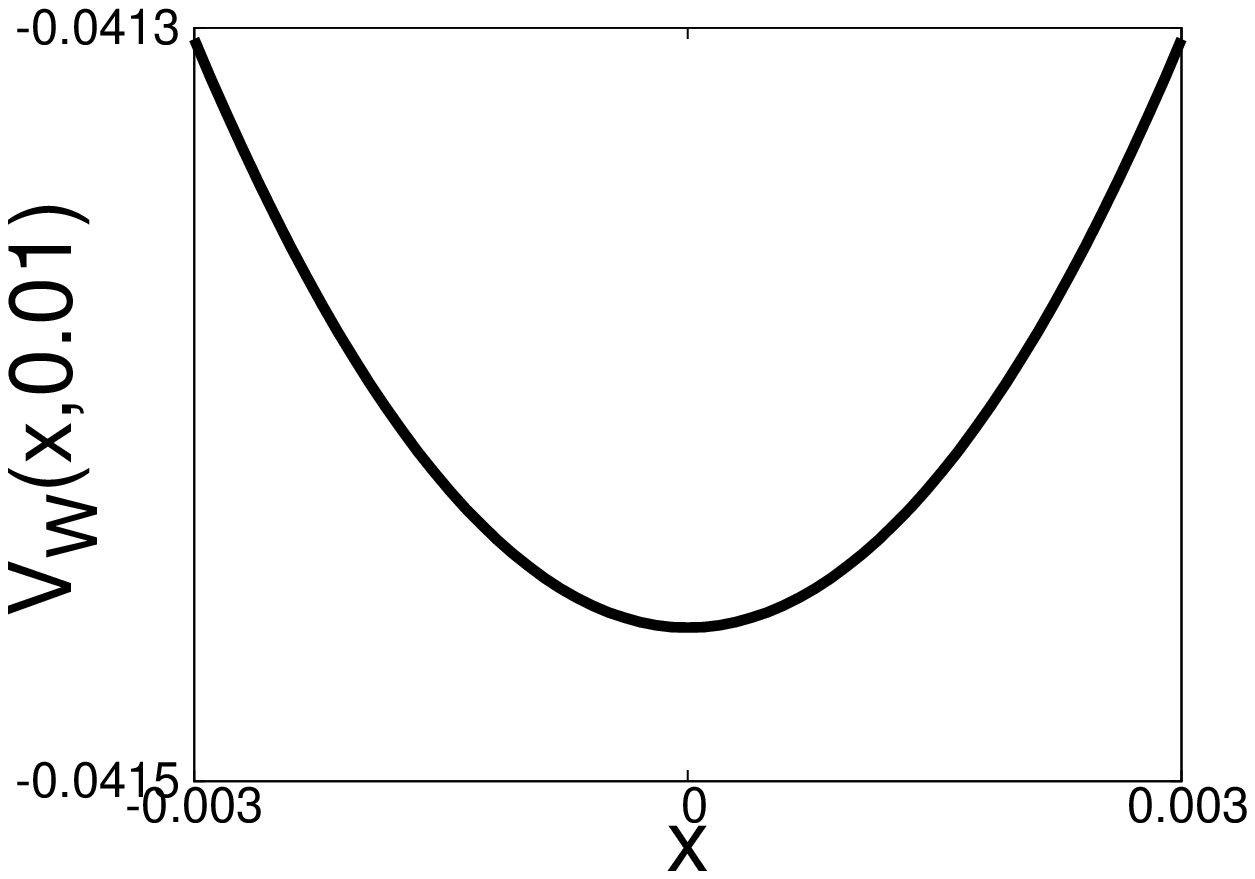}
          \hspace{1cm} (1b) 
        \end{center}
      \end{minipage}
      \begin{minipage}{0.3\hsize}
        \begin{center}
          \includegraphics[clip, width=5cm]{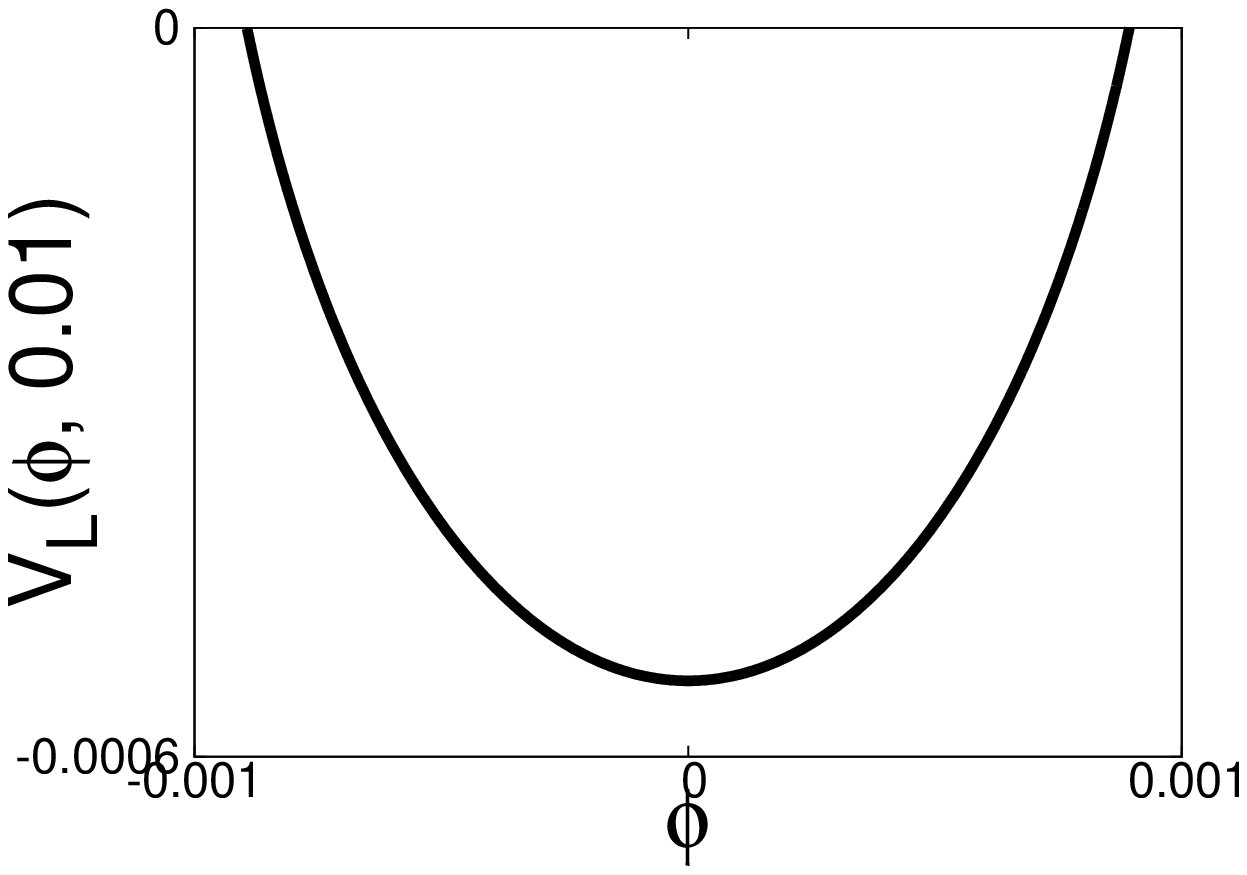}
          \hspace{1cm} (1c) 
        \end{center}
      \end{minipage}
    \end{tabular}
\vskip 0.3cm
    \begin{tabular}{c}
      \begin{minipage}{0.3\hsize}
        \begin{center}
          \includegraphics[clip, width=5cm]{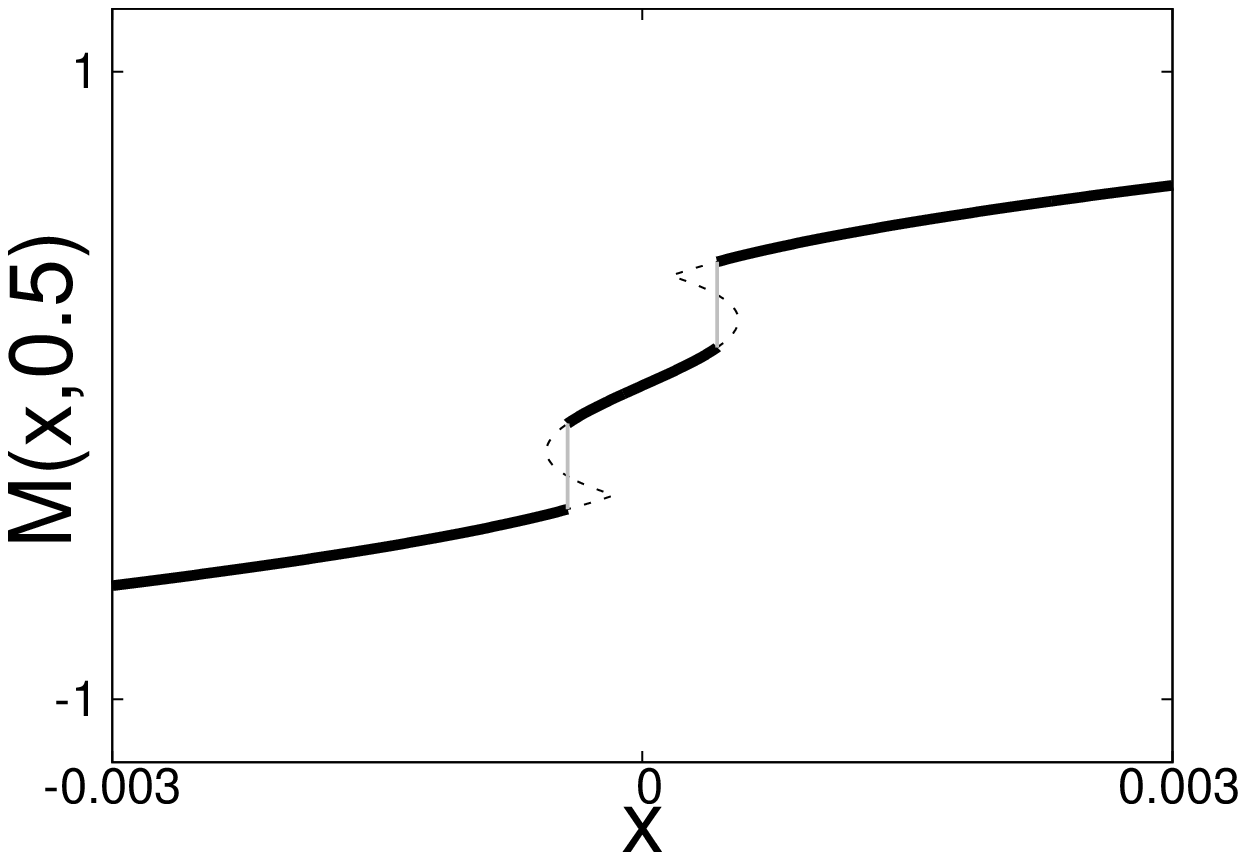}
          \hspace{1cm} (2a) 
        \end{center}
      \end{minipage}
      \begin{minipage}{0.3\hsize}
        \begin{center}
          \includegraphics[clip, width=5cm]{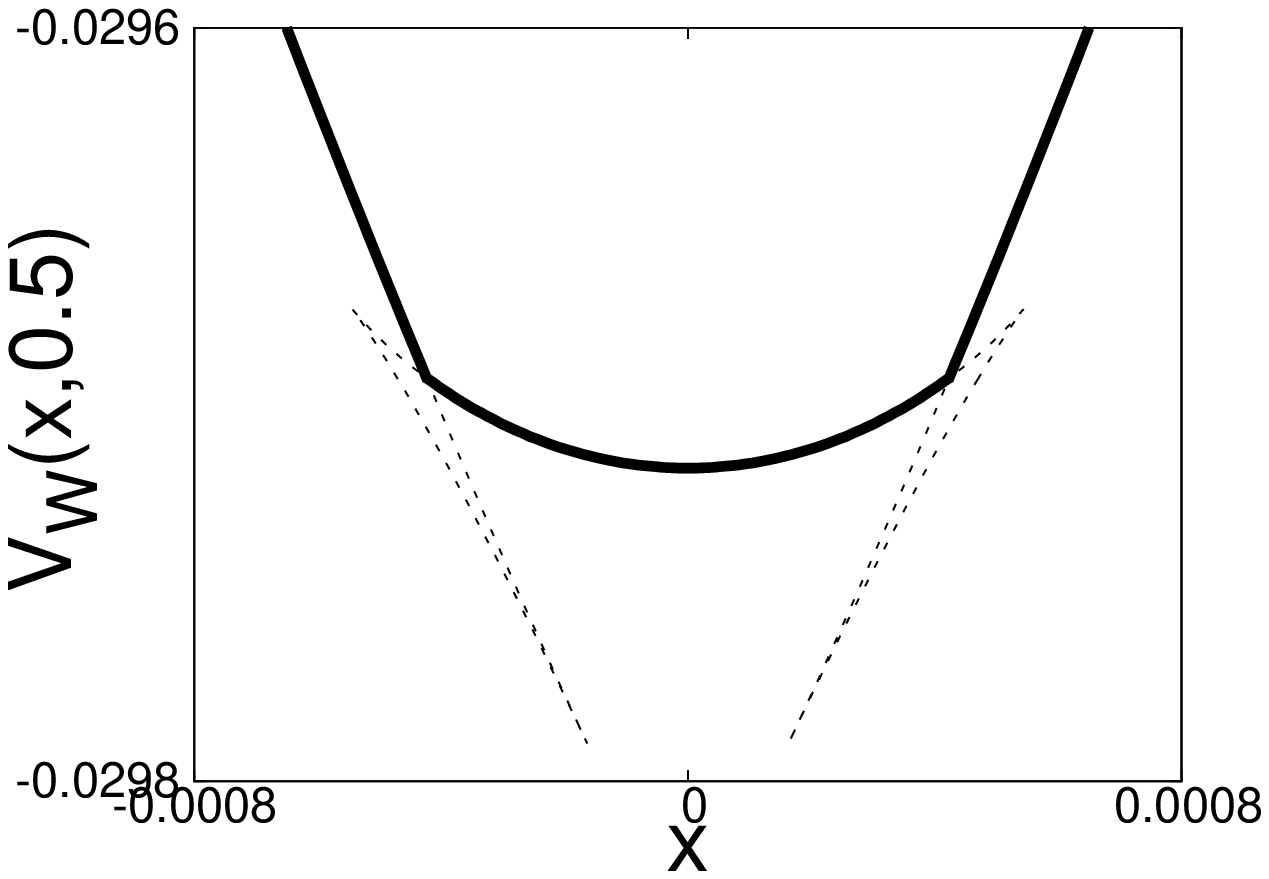}
          \hspace{1cm} (2b) 
        \end{center}
      \end{minipage}
      \begin{minipage}{0.3\hsize}
        \begin{center}
          \includegraphics[clip, width=5cm]{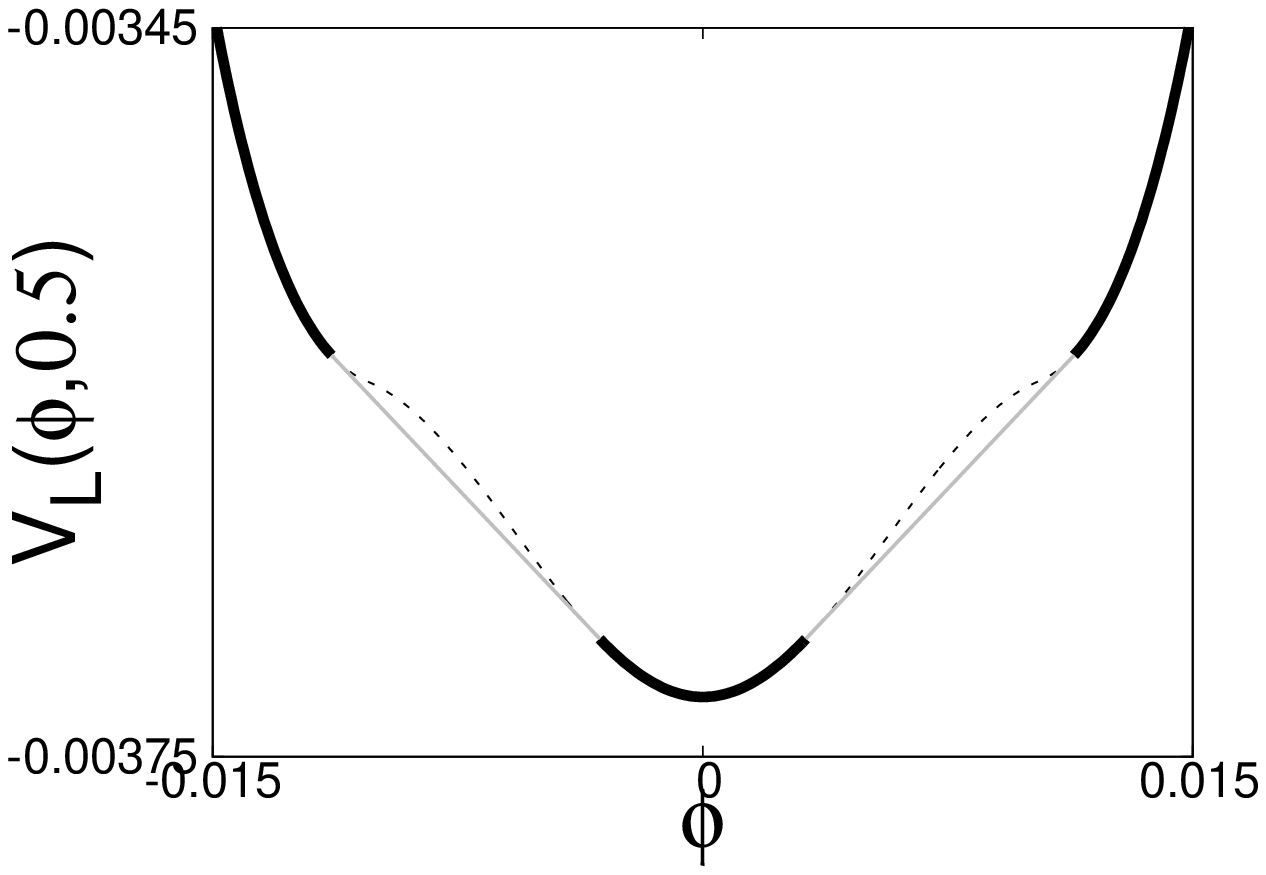}
          \hspace{1cm} (2c) 
        \end{center}
      \end{minipage}
    \end{tabular}      
\vskip 0.3cm
    \begin{tabular}{c}
      \begin{minipage}{0.3\hsize}
        \begin{center}
          \includegraphics[clip, width=5cm]{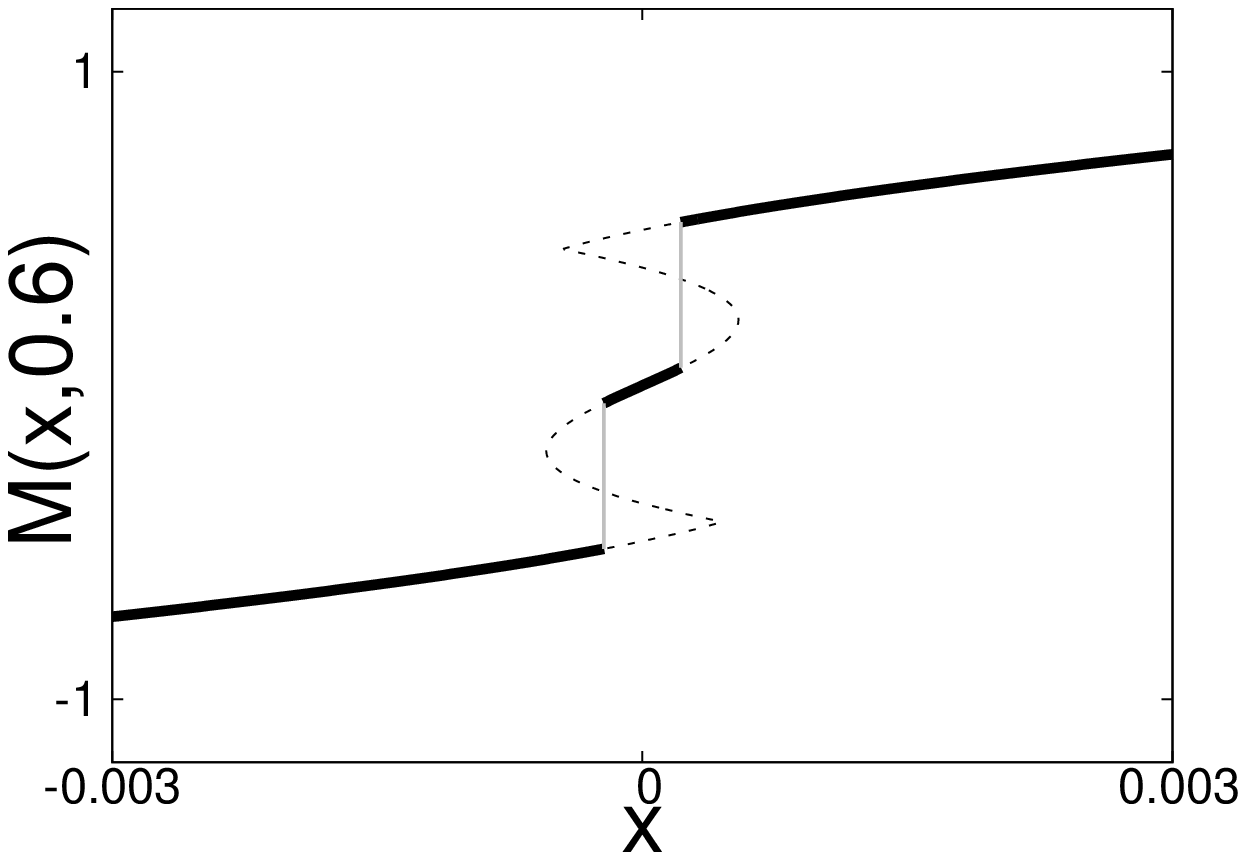}
          \hspace{1cm} (3a) 
        \end{center}
      \end{minipage}
      \begin{minipage}{0.3\hsize}
        \begin{center}
          \includegraphics[clip, width=5cm]{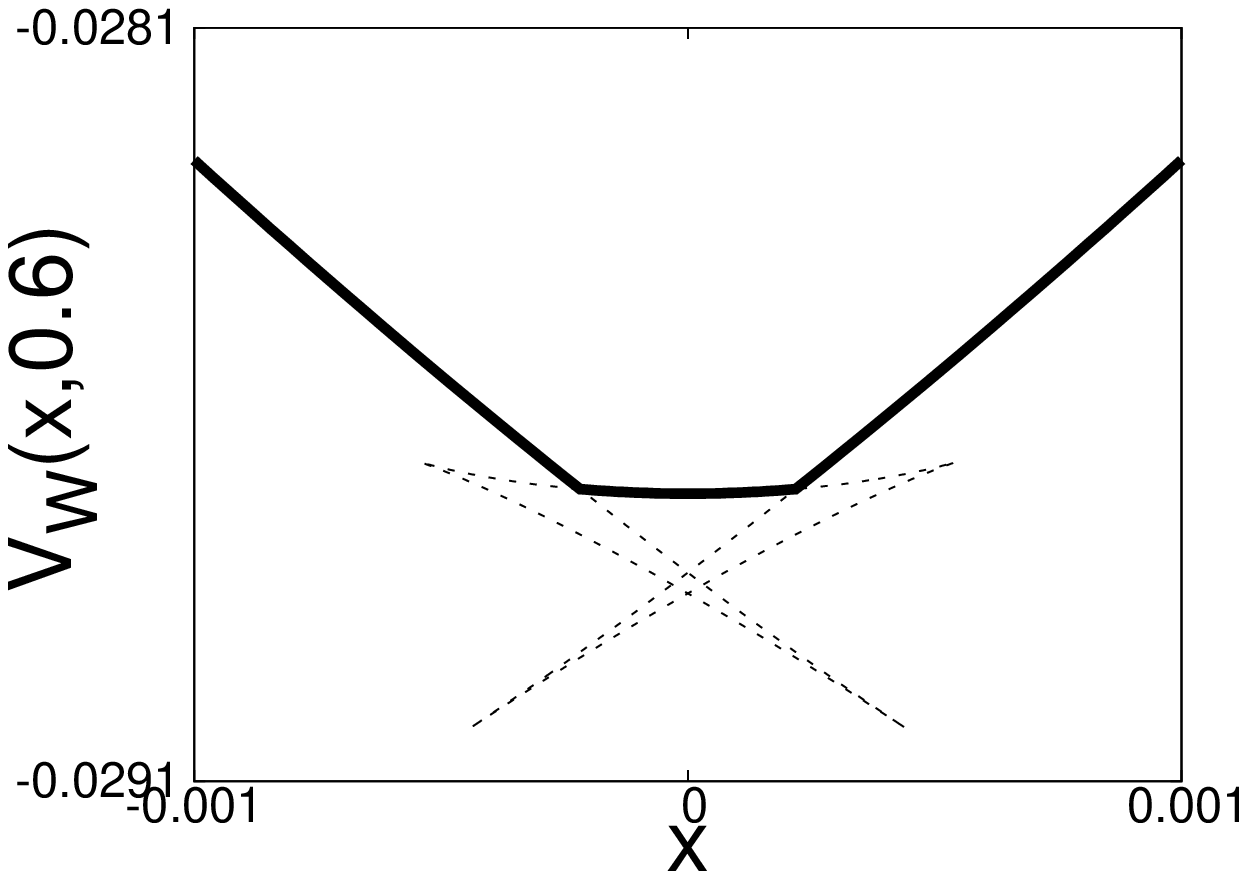}
          \hspace{1cm} (3b) 
        \end{center}
      \end{minipage}
      \begin{minipage}{0.3\hsize}
        \begin{center}
          \includegraphics[clip, width=5cm]{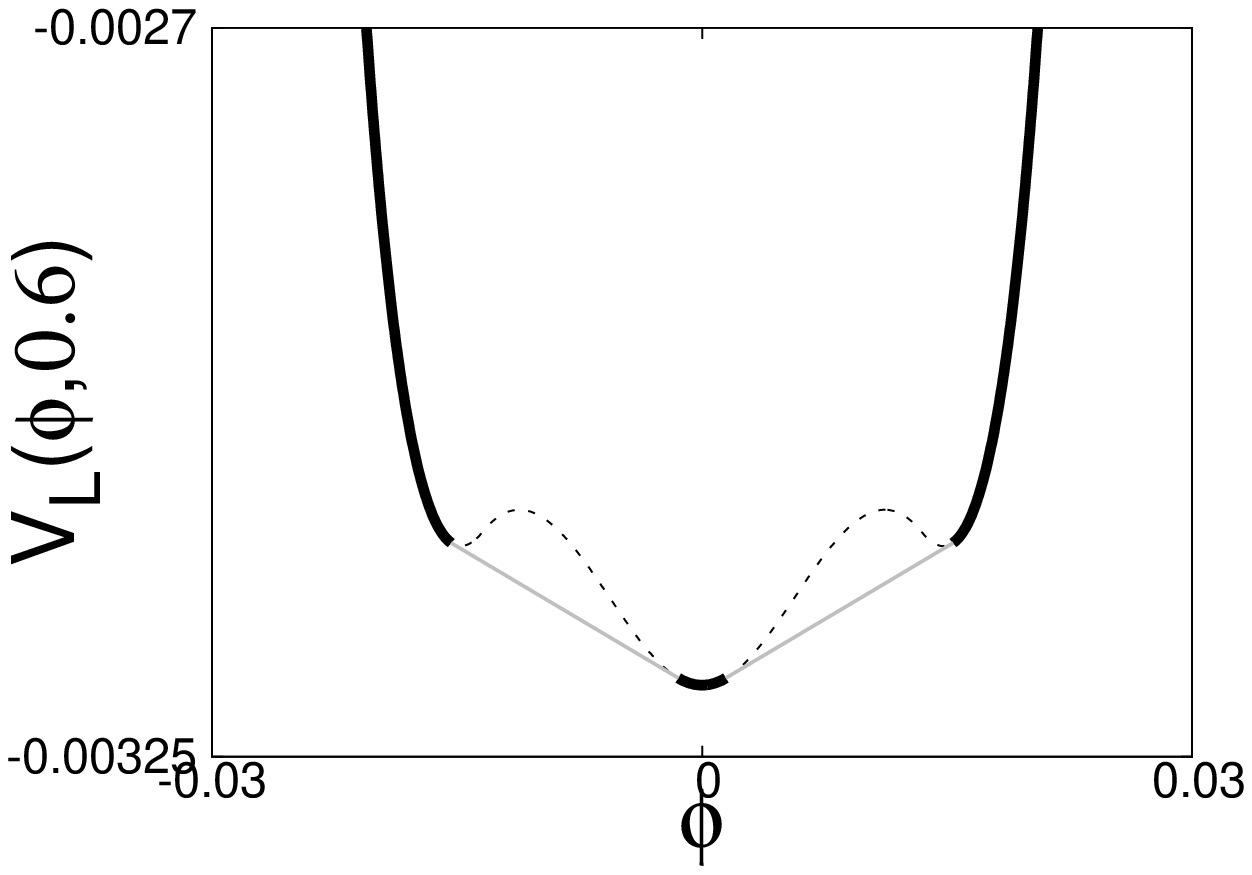}
          \hspace{1cm} (3c) 
        \end{center}
      \end{minipage}
    \end{tabular}
\vskip 0.3cm
    \begin{tabular}{c}
      \begin{minipage}{0.3\hsize}
        \begin{center}
          \includegraphics[clip, width=5cm]{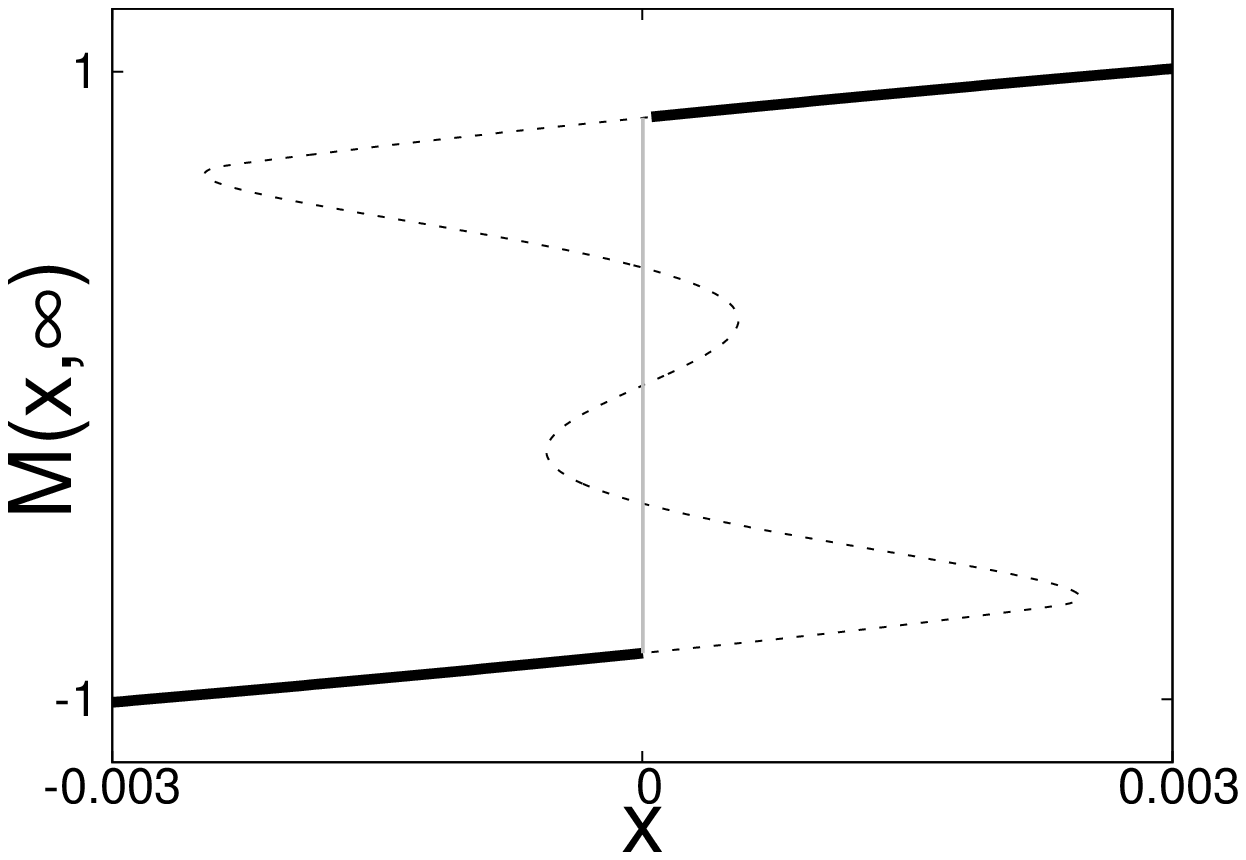}
          \hspace{1cm} (4a) 
        \end{center}
      \end{minipage}
      \begin{minipage}{0.3\hsize}
        \begin{center}
          \includegraphics[clip, width=5cm]{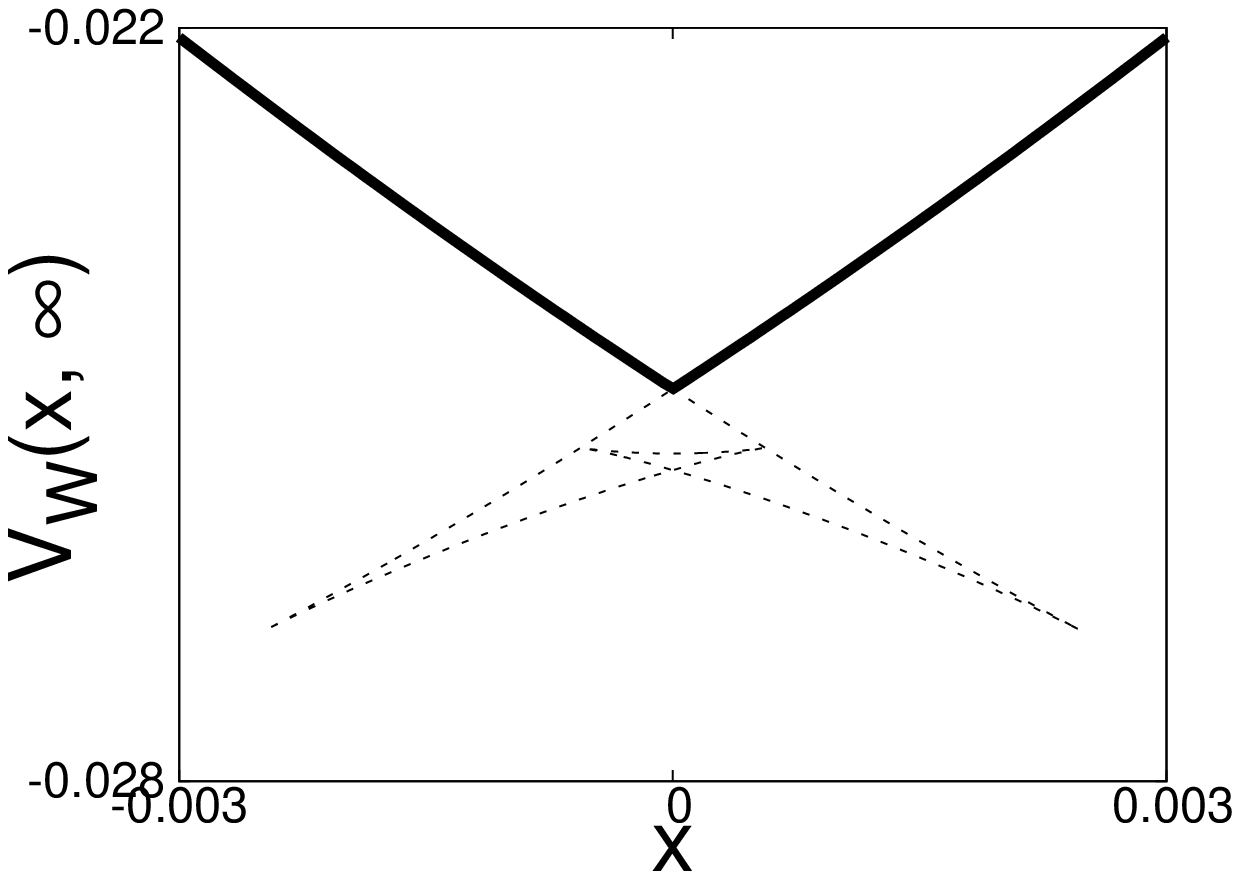}
          \hspace{1cm} (4b) 
        \end{center}
      \end{minipage}
      \begin{minipage}{0.3\hsize}
        \begin{center}
          \includegraphics[clip, width=5cm]{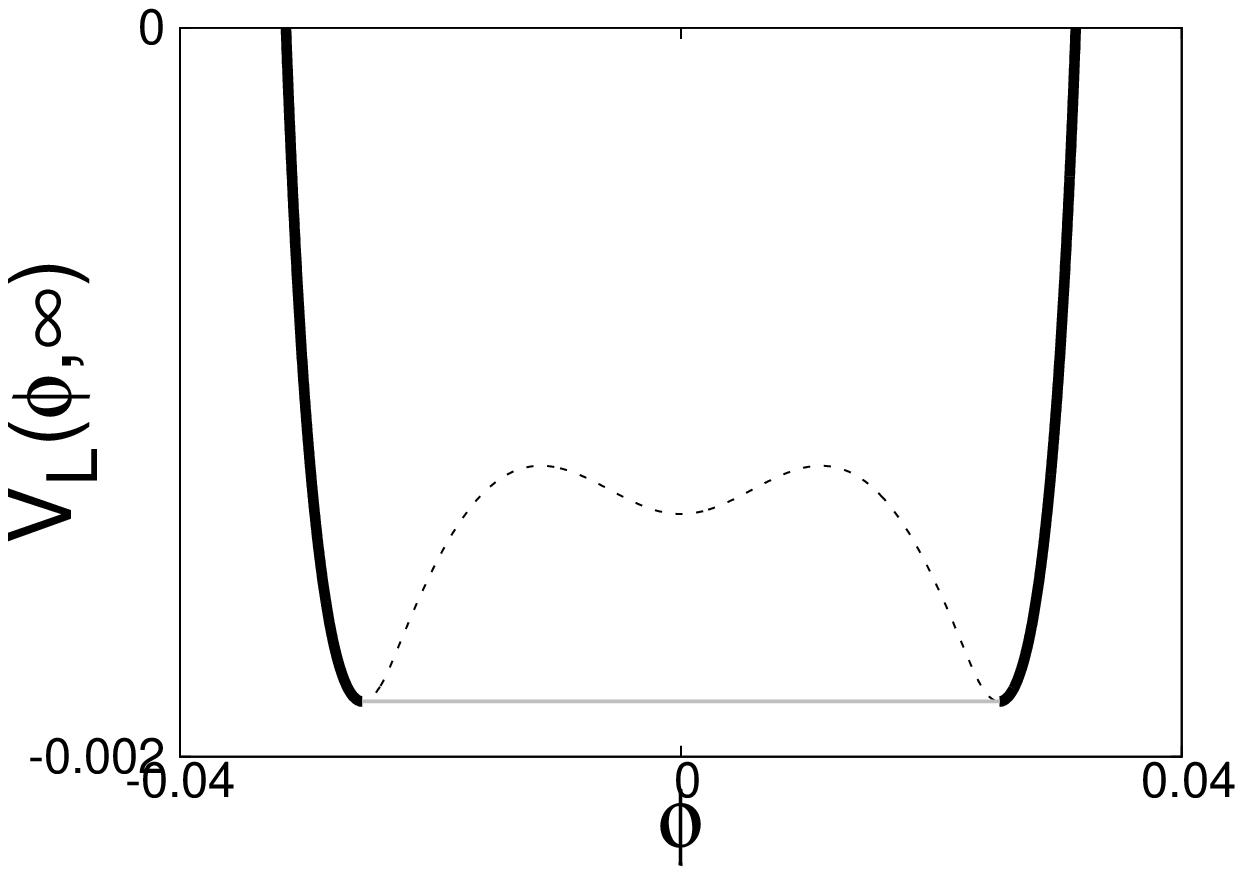}
          \hspace{1cm} (4d) 
        \end{center}
      \end{minipage}     
    \end{tabular}
    \caption{Evolution of physical quantities by weak solution 
(NJL $g=1.7 g_{\rm c}$,$~\mu=0.7$,$~t=0.01$, $0.5$, $0.6$, $\infty$). (a) Mass function. (b) Wilsonian fermion potential. (c) Legendre effective potential.}    
    
    \label{fig:fig3}
  \end{center}
\end{figure}

\clearpage
\begin{thebibliography}{9}
\bibitem{Aoki96} K-I.~Aoki, {\em Proc. SCGT96}, 171 (1996):hep-ph/9706264,
{\em Prog. Theor. Phys. Suppl.} {\bf 131}, 129 (1998), 
{\em Int. J. Mod. Phys. B} {\bf 14}, 1249 (2000).
\bibitem{Aoki99} K-I.~Aoki, K.~Morikawa, J.-I.~Sumi, H.~Terao and M.~Tomoyose, 
{\em Prog. Theor. Phys.} {\bf 102}, 1151 (1999), {\em Phys. Rev. D} {\bf 61}, 045008 (2000). 
\bibitem{Gies02} H.~Gies and C. Wetterich {\em Phys. Rev. D} {\bf 65}, 065001 (2002).
\bibitem{Miya09} K-I.~Aoki and K.~Miyashita {\em Prog. Theor. Phys.} {\bf 121}, 875 (2009).
\bibitem{Eva01} L.~C.~Evans, {\em Partial Differential Equations}, 2nd ed. (AMS, 2010).
\bibitem{Sato13} K-I.~Aoki and D.~Sato {\em Prog. Theor. Exp. Phys.} {\bf  2013}, 043B04 (2013). 
\bibitem{Kuma13} K-I.~Aoki and S.-I.~Kumamoto and D.~Sato in preparation.    
\end{thebibliography}
\end{document}